# Multimode piezoelectric shunt damping of thin plates with arrays of separately shunted patches, method, and experimental validation

**Peyman Lahe Motlagh[a,b*], Ipek Basdogan[a]**

## Abstract

Two-dimensional thin plates are widely used in many applications. Shunt damping is a promising way for the attenuation of vibration of these electromechanical systems. It enables a compact vibration damping method without adding significant mass and volumetric occupancy. Analyzing the dynamics of such electromechanical systems requires precise modeling tools that properly consider the coupling between the piezoelectric elements and the host structure. Although the concept of shunt damping has been studied extensively in the literature, most of the studies do not provide a formulation for modeling the multiple piezoelectric patches that are scattered on the host structure and shunted separately. This paper presents a methodology and a formulation for separately shunted piezoelectric patches for achieving higher performance on vibration attenuation. The Rayleigh-Ritz method is used for performing modal analysis and obtaining the frequency response functions of the electro-mechanical system. The developed model includes mass and stiffness contribution of the piezoelectric patches as well as the electromechanical coupling effect. In this study, the piezoelectric patches are shunted via separate electrical circuits and compared with the ones those are shunted via interconnected electrical circuits. For verification, system-level finite element simulations are performed in ANSYS software and compared with the analytical model results. An experimental setup is also built to validate the performance of the separately shunted piezoelectric patches. The effectiveness of the method is investigated for a broader range of frequencies and it was shown that separately shunted piezoelectric patches are more effective compared to connected for a wide range of frequencies.

[a]**Department of Mechanical Engineering, College of Engineering, Koc University, Istanbul, Turkey**
[b]**Faculty of Engineering and Natural Sciences, Sabanci University, Istanbul 34956, Turkey**
*****Corresponding Author:**
**Peyman Lahe Motlagh, Faculty of Engineering and Natural Sciences, Sabanci University, Istanbul 34956, Turkey**
**e-mail: peyman.lahe@sabanciuniv.edu**

# 1 Introduction

Mechanical structures are exposed to vibrations caused by operational or environmental sources, which can be undesirable, Over the decades, many passive and semi-passive systems have been proposed for reducing these vibrations [1,2]. Piezoelectric structures have been widely used in a range of applications including vibration control [2,3], energy harvesting [1,4], structural health monitoring [5,6]. Over the past few decades, among the transducers that convert the mechanical (electrical) energy to electrical (mechanical) energy, piezoelectric transducers are mostly preferred to the electromagnetic [7,8], and electrostatic [7,9], ones due to their high power density and ease of manufacturing at different size scales [10]. The most common use of piezoelectric materials in the form of patches/layers is by integrating them to the surfaces of flexible beam/plate-like structures, and then utilize them in bending motion for generating an electrical signal and vice versa (applying a voltage to generate a bending deformation).

Piezoelectric patches have been used for shunt damping applications that focus on designing simple electrical circuits that efficiently reduce the structural vibrations [11,12]. Ideally, performance requirements include stability and low energy consumption. The shunt circuit is said to be passive if it does not require an external power supply (e.g. R-shunt) and semi-passive if the circuit operation needs external power supply but does not deliver any power to the mechanical structure (e.g. SSDI) [13]. A resistor connected to the piezoelectric transducer provides the simplest means of achieving energy dissipation and thus vibration damping [14,15]. Another type of circuit that can be used is formed by the integration of a resistor (R) and an inductor (L), which is recognized as an effective method for achieving vibration attenuation. The RL shunt circuit was first introduced in 1979 by Forward [11]. In those cases, the piezoelectric transducer can be tuned to a specific frequency for achieving vibration reduction. Combining the R and L segments in-parallel rather than in-series configuration was first proposed by Wu [16]. One main drawback of this configuration is that RL shunt circuits designed to operate in low-frequency range tend to have large values of inductance. This can be a serious limitation since it would require impractical large and heavy coil inductors. To overcome this difficulty, different methods were developed including the use of electronic circuits based on operational amplifiers, but it still needs an external source of energy to operate [17]. While single-mode shunt circuits can only be tuned to one frequency, multi-mode shunts can be tuned to dampen several structural modes at the same time [18]. The main difficulty with this method is that the circuit order (the number of elements which is used) increases very rapidly with the number of the targeted modes (i.e., the number of necessary inductors is the square of the number of modes). Another method for shunt damping is using a negative capacitance circuit [20, 21]. Negative capacitances cannot be found commercially. However, they can be achieved by using active circuits [21], including, an operational amplifier, a capacitor, and resistors, which make it impractical to implement and also limited usage. A new way of shunt damping is the SSDI circuits [13], in which a small inductance is used to quickly invert the electrical charge at the electrodes. For this purpose, an electrical switch is utilized to adjust the maximum deformation of the piezoelectric transducers to achieve maximum vibration reduction at the desired frequency [14, 21]. However, being optimized to a single frequency is the drawback of those systems which is not very efficient when broadband vibration reduction is desired [22].

In energy harvesting, vibration control, and actuation/sensing applications, piezoelectric materials are typically in the form of thin square patches that are bonded to specific locations on the surface of the thin plates. For the implementation of these structures, Aridogan et.al. extended the electro-elastic beam model [23] to thin plates with structurally integrated single/multiple piezoelectric patch harvesters [24]. They experimentally validated their proposed electro-elastic model which could accurately predict the electromechanical behavior by only including the induced moments and neglecting the inertial and stiffness contributions of the piezoelectric-patches. Later on, Motlagh et. al. extend works of [25] to achieve modeling of curved structures [26,27] and functionally graded materials [4].

Based on the above studies conducted on the modeling of piezoelectric patches bonded on thin plate structures, a new approach is proposed in this study to obtain broadband frequency shunt damping. The Raleigh Ritz model developed in [30] is extended to cover the electro-mechanical equations for separately shunted circuits and also neutral axis shift is included in the electro-mechanical model to accommodate when a single patch is used on one side of the host plate. The results of the extended electro-mechanical model are validated by FEM results and experiments are conducted to demonstrate the accuracy of the proposed technique of the separately shunted circuits. In this study, it is shown that the distribution of the optimized resistor values of the shunt circuits improves the shunt damping performance which is a significant observation of the present study.

## 2 Analytical model of a thin plate with multiple piezoelectric patches

In this section, a brief description of the model of a thin plate with multiple piezoelectric patches is given based on the Kirchhoff plate theory [28]. Figure 1, presents the host plate with all four edges clamped (CCCC) boundary conditions and the structurally integrated piezoelectric patches in separated and connected configurations, respectively. As it can be seen from the figure, in the separated configuration, each patch is connected independently to a shunted electric circuit and the negative end is grounded on the plate, whereas in the connected configuration, all the positive ends of the patches are connected to the positive end of each shunted electric circuit and the negative ends are grounded on the plate.

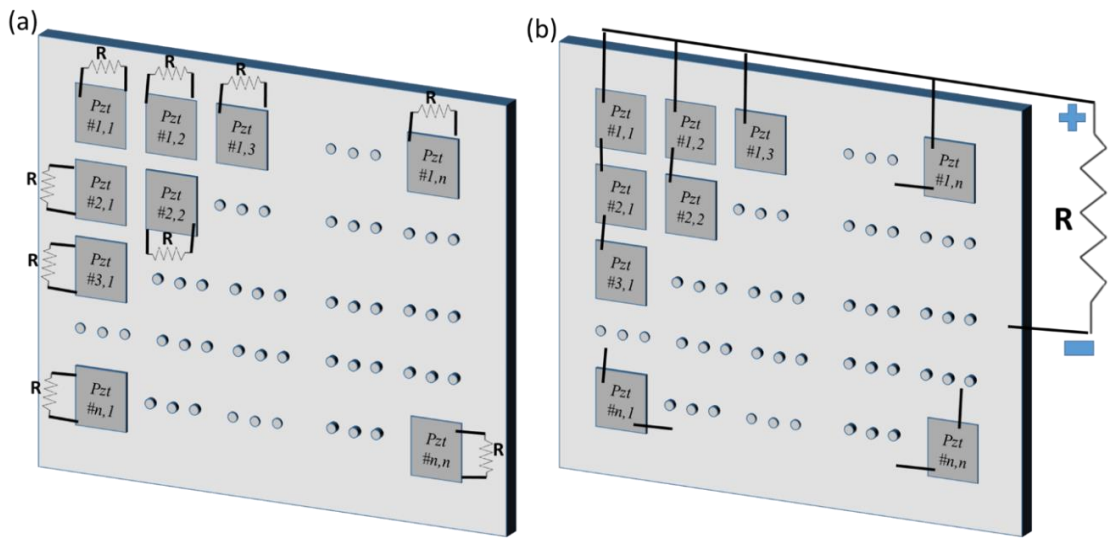

**Figure 1. (a) separated configuration of the piezoelectric patches and the host plate, (b) connected configuration of the piezoelectric patches and the host plate**

## 3 Constitutive equations of the thin plate with piezoelectric patches based on Kirchhoff plate theory

Since the piezoelectric patches are typically manufactured as a thin plate, piezoelectric patch skin can be modeled as a two-dimensional Kirchhoff plate. According to Kirchhoff plate theory [28], the deflection of the middle surface is assumed to be small compared to the thickness of the skin. Since the normal stress $\sigma_{zz}$ in the thickness direction is much smaller compared to the in-plane stresses, it can be ignored under the assumption of the thin plate theory. The material of the structural layer is assumed to be isotropic. Therefore, based on the Kirchhoff plate theory, the constitutive equations for the structural layer are reduced to Eq. (1) where $Y_s$ is the Young's modulus and $v_s$ is the Poisson's ratio of the structural layer. The in-plane stress and strain components for the piezoelectric patches are denoted by $\sigma_{ij}$ and $\gamma_{ij}, \varepsilon_{ij}$ respectively, and

$$\begin{pmatrix} \sigma_{xx} \\ \sigma_{yy} \\ \tau_{xy} \end{pmatrix} = \frac{Y_s}{1-v_s^2} \begin{bmatrix} 1 & v_s & 0 \\ v_s & 1 & 0 \\ 0 & 0 & (1-v_s)/2 \end{bmatrix} \begin{pmatrix} \varepsilon_{xx} \\ \varepsilon_{yy} \\ \gamma_{xy} \end{pmatrix} \quad (1)$$

The constitutive equations of a piezoelectric patch are expressed in a reduced form as Eq. (2) [28]:

$$\begin{pmatrix} \sigma_{xx} \\ \sigma_{yy} \\ \tau_{xy} \\ D_3 \end{pmatrix} = \begin{bmatrix} \bar{c}_{11} & \bar{c}_{12} & 0 & -\bar{e}_{31} \\ \bar{c}_{12} & \bar{c}_{11} & 0 & -\bar{e}_{31} \\ 0 & 0 & \bar{c}_{66} & 0 \\ \bar{e}_{31} & \bar{e}_{31} & 0 & \varepsilon^S_{33} \end{bmatrix} \begin{pmatrix} \varepsilon_{xx} \\ \varepsilon_{yy} \\ \gamma_{xy} \\ E_{33} \end{pmatrix} \quad (2)$$

Where $\bar{c}_{ij}$ are reduced elastic moduli of the piezoelectric patches, $\bar{e}_{ij}$ and $\varepsilon^S_{33}$ are piezoelectric constant and dielectric permittivity, respectively [28]. Hamilton's principle is used to determine the equation of motion Eq. (3):

$$\delta \oint_{t_1}^{t_1} (KE - PE + W_p) \, dt = 0 \quad (3)$$

Where kinetic energy, potential energy, and applied external work is indicated as $KE$, $PE$, and $W_p$ respectively [29]. And the indicator function $P(x, y)$ is defined to identify the areas where $k$ piezoelectric patches are attached to the surface of the structural layer by Eq. (4):

$$P(x,y) = \sum_{i=1}^{k} [H(x - x_{i,1}) - H(x - x_{i,2})] \times [H(y - y_{i,1}) - H(y - y_{i,2})] \quad (4)$$

Where $x_1, x_2, y_1, and\ y_2$ are the vertices of the area along x- and y-axes, respectively, and H denotes the Heaviside unit step function. The equivalent mass per unit area can be derived as Eq. (5):

$$m(x,y) = \rho_s h_s + \rho_p h_p P(x,y) \quad (5)$$

Here $\rho_s$, $\rho_p$, are densities and $h_s$ and $h_p$ are the thicknesses of the host plate and the piezoelectric patch, respectively. The stress distribution in the areas covered by the piezoelectric patch is not the same as that in the uncovered areas. Thus, the neutral surface is shifted from the middle surface of the structural layer towards the piezoelectric patch. Based on the stress equilibrium, the location of the neutral surface can be determined as Eq. (6):

$$z_0 = \frac{\bar{c}_{11} h_p (h_s + h_p)}{2 \left\{ \frac{Y_s h_s}{(1-v_s^2)} + \bar{c}_{11} h_p \right\}} \quad (6)$$

Then, the equation of motion of the plate and piezoelectric patches can be derived as Eq. (7) in which $w$ is displacement of the system:

$$m(x,y)\ddot{w} + \left[\{D^s + P(x,y)D^{sp}\}\left\{\begin{array}{c}\left(\frac{\partial^2 w}{\partial x^2}\right)^2 + \left(\frac{\partial^2 w}{\partial y^2}\right)^2 + 2v_s\left(\frac{\partial^2 w}{\partial x^2}\frac{\partial^2 w}{\partial y^2}\right) \\ +2(1-v_s)\left(\frac{\partial^2 w}{\partial x \partial y}\right)^2\end{array}\right\}\right.$$

$$+ P(x,y)D_{11}{}^p\left\{\left(\frac{\partial^2 w}{\partial x^2}\right)^2 + \left(\frac{\partial^2 w}{\partial x^2}\right)^2\right\}$$

$$\left. + 2P(x,y)\left\{D_{12}{}^p\left(\frac{\partial^2 w}{\partial x^2}\frac{\partial^2 w}{\partial y^2}\right) + 2D_{66}{}^p\left(\frac{\partial^2 w}{\partial x \partial y}\right)^2\right\}\right]$$

$$+ P(x,y)\bar{e}_{31}v(t)\left(\frac{h_s + h_p}{2} - z_0\right)\left(\frac{\partial^2 w}{\partial x^2} + \frac{\partial^2 w}{\partial y^2}\right)$$

$$= f(t)\delta(x-x0)\delta(y-y0)$$

(7)

where $D_{11}{}^p, D_{12}{}^p, D_{66}{}^p, D^s$ and $D^{sp}$ are as follows

$$D_{11}{}^p = \bar{c}_{11}\left(\frac{h_p{}^3}{3} + \frac{h_s{}^2 h_p}{4} + \frac{h_s h_p{}^2}{2} - z_0(h_p h_s + h_p{}^2) + z_0{}^2 h_p\right)$$

$$D_{12}{}^p = \bar{c}_{12}\left(\frac{h_p{}^3}{3} + \frac{h_s{}^2 h_p}{4} + \frac{h_s h_p{}^2}{2} - z_0(h_p h_s + h_p{}^2) + z_0{}^2 h_p\right) \quad (8)$$

$$D_{66}{}^p = \bar{c}_{66}\left(\frac{h_p{}^3}{3} + \frac{h_s{}^2 h_p}{4} + \frac{h_s h_p{}^2}{2} - z_0(h_p h_s + h_p{}^2) + z_0{}^2 h_p\right)$$

$$D^s = \frac{Y_s h_s{}^3}{12(1-v_s{}^2)}$$

$$D^{sp} = \frac{Y_s}{1-v_s{}^2}\left(\frac{h_s{}^3}{12} + z_0{}^2 h_p\right)$$

(9)

### 3.1 Modal analysis using the Rayleigh-Ritz method

The natural frequencies and corresponding mode shapes are calculated for the fully clamped (CCCC) boundary conditions. Based on the modal expansion, the relative displacement of piezoelectric patch skin is approximated by a linear combination of the assumed modes $U_{ij}W_{ij}(x,y)$ as Eq. (10), where $\mu_{ij}(t)$ are the generalized modal coordinates, N is the total number of vibration modes in y and R is the total number of vibration modes x coordinates. Assumed modes are indicated by $U_{ij}W_{ij}(x,y)$, where $W_{ij}(x,y)$ are the trial functions satisfying the boundary conditions, and $U_{ij}$ are the corresponding coefficients.

$$w(x,y,t) = \sum_{i=1}^{N}\sum_{j=1}^{N} U_{ij} W_{ij}(x,y)\mu_{ij}(t) \quad (10)$$

The eigenvalue equation can be written as Eq. (11) for more information, the reader can be referred to [28]:

$$[K_{rn,kl} - \omega_{rn}^2 M_{rn,kl}][U_{rn}] = \{0\} \quad (11)$$

Here, assumed mode shape coefficients $U_{ij}$'s are the eigenvectors and natural frequencies $\omega_{ij}$'s are the square root of the eigenvalues of Eq. (11).

### 3.2 Electromechanical coupling between the patches and the host structure

The electric current can be calculated using Gauss's law (i.e., Gauss's flux theorem) with respect to the electric displacement component $D_3$. This value is the same as the scalar product of the electric displacement field D and the outward normal unit vector n [30] as Eq. (12) in which $v(t)$, $R_l$ and $S_p$ are voltage over the piezoelectric patch, resistor attached to each patch and area of piezoelectric patch, respectively.

$$\frac{v(t)}{R_l} = \frac{d}{dt}\int_{S_p}(D.n)dS_p = \frac{d}{dt}\int_{S_p}D_3 dS_p \qquad (12)$$

By substituting $D_3$ from Eq. (2) in Eq. (12) and substituting strain Eq. (2), Eq. (12) can be written as Eq. (13).

$$\frac{v(t)}{R_l} = -\bar{e}_{31}\left(\frac{h_p+h_s}{2}-z_0\right)\frac{\partial}{\partial t}\int_{S_p}\nabla^2 w(x,y,t)dS_p - \frac{dv(t)}{dt}\int_{S_p}\frac{\varepsilon^S_{33}}{h_p}dS_p \qquad (13)$$

The last term on the right-hand side of Eq. (13) can be considered as the capacitance $C_p^k$ of the $k^{th}$ piezoelectric patch which can be defined as Eq. (14):

$$C_p^k = \int_{S_p}\frac{\varepsilon^S_{33}}{h_p}dS_p \qquad (14)$$

The Kirchhoff laws can be applied to the electrical circuits to obtain Eq. (15), by comparing Eq. (13) and Eq. (15), the current passing through $k^{th}$ piezoelectric patch is defined as Eq. (16) [30].

$$C_p^k\left(\frac{dv(t)}{dt}\right) + \frac{v(t)}{Z_l} = i_k(t) \qquad (15)$$

$$i_k(t) = -\bar{e}_{31}\left(\frac{h_p+h_s}{2}-z_0\right)\frac{\partial}{\partial t}\int_{S_p}\nabla^2 w(x,y,t)dS_p \qquad (16)$$

Then, the current output in modal coordinates can be obtained by substituting Eq. (10) into Eq. (16), and written as:

$$i_k(t) = -\sum_{n=1}^{N}\sum_{r=1}^{R}\frac{d\eta_{rn}(t)}{dt}\tilde{\theta}_{rn}^{k} \qquad (17)$$

Where $\tilde{\theta}_{rn}^{k}$ can be defined as Eq. (17):

$$\tilde{\theta}_{rn}^{k} = -\bar{e}_{31}\left(\frac{h_p+h_s}{2}-z_0\right)\int_{S_p}U_{rn}\nabla^2 W_{rn}(x,y)dS_p \qquad (18)$$

### 3.3 Steady-state response

The steady-state equations for piezo patches on a plate can be divided into two categories, connected and separated configurations. The connected patches were previously presented in literature [4, 5, 31]. In this section, first, connected configuration equations are summarized then the separated configuration equations will be derived. For each of these cases, the current through each of piezoelectric patch can be represented using Eq. (17) [24]. If the transverse force excitation on the host plate is assumed to be harmonic of the form $f(t) = F_0 e^{j\omega t}$ (where $F_0$ is the force amplitude and $\omega$ is the excitation frequency), the linear system assumption implies that the mechanical and electrical responses can be represented as harmonic functions: $\eta_{rn}(t) = H_{rn} e^{j\omega t}$ and $v_k(t) = V_k e^{j\omega t}$, respectively.

### 3.3.1 Connected patches

In the connected configuration, positive nodes of all of the patches are connected to each other whereas all of their negative nodes are considered grounded ($v = 0$) as in Figure 1. For this configuration, the circuit equation can be written as Eq. (19) using Kirchhoff's current law, where $Z_l$ is impedance of load attached to each piezoelectric patch [24]:

$$\sum_{k=1}^{n} C_p^k \left( \frac{dv(t)}{dt} \right) + \frac{v(t)}{Z_l} = \sum_{k=1}^{n} i_k(t) \tag{19}$$

A complete methodology description of connected patches is presented in [24], in which the displacement of the system $w(x, y, t)$ is given by:

$$w(x, y, t) = \sum_{r=1}^{R} \sum_{n=1}^{N} U_{rn} W_{rn}(x, y) \left\{ \frac{F_0 U_{rn} W_{rn}(x_0, y_0) + V_k \sum_{k=1}^{K} \tilde{\theta}_{rn}^k}{\omega_{rn}^2 - \omega^2 + 2j\xi_{rn}\omega_{rn}\omega} \right\} e^{j\omega t} \tag{20}$$

### 3.3.2 Separated patches

In the separated configuration, each patch is connected to an electric circuit independently as in Figure 1-(a), by applying Kirchhoff's current law for each patch, the circuit equation can be written as Eq. (21):

$$C_p^k \left( \frac{dv_k(t)}{dt} \right) + \frac{v_k(t)}{(Z_l)_k} = i_k(t), (k = 1, 2, \dots, Number\ of\ patches) \tag{21}$$

The independent differential equations of motion in modal coordinates was developed previously in [30], [28] as Eq. (22) where the voltage $v_k(t)$ will be different on each patch:

$$\frac{d^2 \eta_{rn}(t)}{dt^2} + 2\omega_{rn}\xi_{rn} \frac{d\eta_{rn}(t)}{dt} + \omega_{rn}^2 \mu_{rn}(t) - \sum_{k=1}^{K} \theta_{rn}^k v_k(t) = f_{rn}(t) \tag{22}$$

Where $f_{rn}(t)$ [28] is :

$$f_{rn}(t) = f(t)\, U_{rn} W_{rn}(x_0, y_0) \tag{23}$$

From Eq. (17) and Eq. (21), the relationship between $W_{rn}$ and $v_k(t)$ can be written as Eq. (24):

$$\left[ \frac{1}{(Z_l)_k} + j\omega C_p^k \right] V_k(t) + j\omega \sum_{n=1}^{N} \sum_{r=1}^{N} \frac{F_0 U_{ij} W_{ij}(x_0, y_0) \tilde{\theta}_{ij}^k}{\omega_{ij}^2 - \omega^2 + (2j)\xi_{ij}\omega_{ij}\omega} \\ + j\omega \sum_{i=1}^{N} \sum_{j=1}^{N} \tilde{\theta}_{ij}^k \frac{\sum_{k=1}^{k} \tilde{\theta}_{ij}^k v_k(t)}{\omega_{ij}^2 - \omega^2 + (2j)\xi_{ij}\omega_{ij}\omega} = 0 \tag{24}$$

Then $V_k(t)$ are the only unknowns in this equation, and they can be derived as:

$$\begin{bmatrix} V_1 \\ \vdots \\ V_k \\ \vdots \\ V_n \end{bmatrix} = A^{-1} \begin{bmatrix} b_1 \\ \vdots \\ b_k \\ \vdots \\ b_n \end{bmatrix} \qquad (25)$$

Where $A$ is defined as Eq. (26):

$$A = \begin{bmatrix} \left\{\frac{1}{(Z_l)_1} + j\omega C_p^{\ 1} + a'_{11}\right\} & \cdots & a'_{1k} & \cdots & a'_{1n} \\ \vdots & \ddots & \ddots & \ddots & \vdots \\ a'_{k1} & \ddots & \left\{\frac{1}{(Z_l)_k} + j\omega C_p^{\ k} + a'_{kk}\right\} & \ddots & a'_{kn} \\ \vdots & \ddots & \ddots & \ddots & \vdots \\ a'_{n1} & \cdots & a'_{nk} & \cdots & \left\{\frac{1}{(Z_l)_n} + j\omega C_p^{\ n} + a'_{nn}\right\} \end{bmatrix} \qquad (26)$$

where $a$ and $b$'s are defined as

$$a'_{ls} = j\omega \sum_{j=1}^{N} \sum_{i=1}^{N} \frac{\tilde{\theta}_{ij}^{\ l} \tilde{\theta}_{ij}^{\ s}}{\omega_{ij}^2 - \omega^2 + 2j\xi_{ij}\omega_{ij}\omega}, (i,j = 1, \ldots, N), (l,s = 1,2,\ldots, \text{number of patches}) \qquad (27)$$

$$b_k = j\omega \sum_{i=1}^{N} \sum_{j=1}^{N} \frac{F_0 U_{ij} W_{ij}(x_0, y_0) \tilde{\theta}_{ij}^{\ k}}{\omega_{rn}^2 - \omega^2 + 2j\xi_{rn}\omega_{rn}\omega}, (k = 1,2,\ldots, \text{number of patches}) \qquad (28)$$

By finding $V = [v_1 \ \cdots \ v_k \ \cdots \ v_n]^T$ from (25), and inserting in (22), the relative displacement of the plate can be defined as Eq. (29):

$$w(x,y,t) = \sum_{i=1}^{N} \sum_{j=1}^{N} U_{ij} W_{ij}(x,y) \left\{ \frac{F_0 U_{ij} W_{ij}(x_0, y_0) + \sum_{k=1}^{K} \theta_{ij} A^{-1}[b_1 \ \cdots \ b_k \ \cdots \ b_n]^T}{\omega_{ij}^2 - \omega^2 + 2j\xi_{ij}\omega_{ij}\omega} \right\} e^{j\omega t} \qquad (29)$$

## 4 Finite element model

Commercial finite element analysis (FEA) software ANSYS is used for validating the modal solutions obtained using the analytical model. The plate is meshed with 20-node structural solid elements (SOLID 186). The piezoelectric patches are meshed with 20-node coupled field solid elements (SOLID 226). The host structure and the piezoelectric patches are bonded by generating constraint equations at the interface [31]. The external resistive load connected to piezoelectric patches is modeled with a piezoelectric circuit element (CIRCU 94). Natural frequencies and mode shapes are obtained by modal analysis.

# 5   Experimental setup

The experimental setup used for comparing the shunt damping performance of the separated and connected patch configurations is presented in Figure 2. A fully clamped aluminum plate is used as the host plate. Three electrode piezoelectric patches (T105-A4E-602 by Piezo Systems, Inc.) are employed for shunt damping by attaching them on the plate. The geometric and material properties of the aluminum plate and the piezoelectric patches are given in Table 1. The three piezoelectric patch harvesters (labeled as PZT-I, PZT-II and PZT-III) having identical polarities are perfectly bonded on the surface of the plate. The bottom (negative) and top (positive) surfaces of the piezoelectric patches are covered by thin vacuum sputtered nickel electrodes of a negligible thickness (by the manufacturer) and the bonding areas on the aluminum plate are electrically insulated with 3M Scotch 1601 spray. A modal shaker is used to excite the plate. The attachment point of the shaker's stinger rod can be seen in Figure 2. The applied force is monitored by placing a force transducer (PCB 208C01) between the shaker's armature and the plate surface. A laser Doppler vibrometer (LDV—Polytec PDV 100) is used to measure the transverse velocity of the plate at the target point shown in Figure 2. The signals obtained from the force transducer and the laser vibrometer are sent to the data acquisition system for time-domain and frequency response analyses.

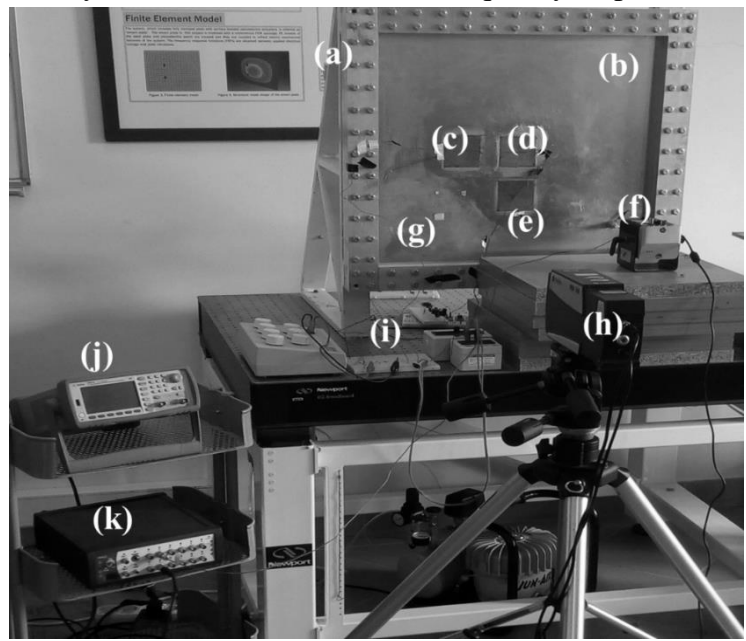

**Figure 2. Experimental setup, (a) clamping frame, (b) aluminum plate, (c) PZT-I, (d) PZT -II, (e) PZT -III, (f) modal shaker, (g) measurement point, (h) laser vibrometer, (i) electrical circuit, (j)signal generator and (k) signal analyzer.**

Table 1. Geometric and material properties of the piezo-patches and the host plate

| Properties | Plate | Piezoelectric (PZT-5A) |
|---|---|---|
| Length (mm) | 540 | 72.4 |
| Width (mm) | 580 | 72.4 |
| Thickness (mm) | 1.9 | 0.267 |
| Young's modulus (GPa) | 70 | 69 |
| Mass density (kg/m3) | 2700 | 7800 |
| Piezoelectric constant (C/m2) | - | -190 |
| Permittivity constant (nF/m) | - | 9.57 |
| Damping ratio | 0.01 | 0.01 |

# Results and discussions

In this section, Experimental validation, Modal analysis comparison with Ansys, and Electromechanical frequency response functions (FRF) results for connected and separated configurations will be presented.

## 5.1 Experimental validations

In order to validate the analytical model presented in the previous section, the FRFs of the system are obtained using the experimental set-up in Section 5. The velocity FRFs are compared for the separated and connected configurations in Figure 2. For both cases, optimum R ($R_{opt}$) is obtained by the analytical model by sweeping the resistor values (R) from short circuit ($R_{SC}$ =100Ω) to open circuit ($R_{OC}$=1 MΩ), then picking the best R that achieves the maximum shunt damping performance. Once the $R_{opt}$ is obtained from the model, it is also implemented in the experimental setup and then compared with the OC (open circuit) configuration. Figures 3a and 3b compare the results for connected and separated configurations, respectively. Please note that, although the $R_{opt}$ values are obtained for the first mode, reductions in the second and third modes can be also observed. The velocity FRFs in the second column of Figure 4, demonstrate the zoomed section for the 2$^{nd}$ and 3$^{rd}$ modes to observe the reductions more clearly.

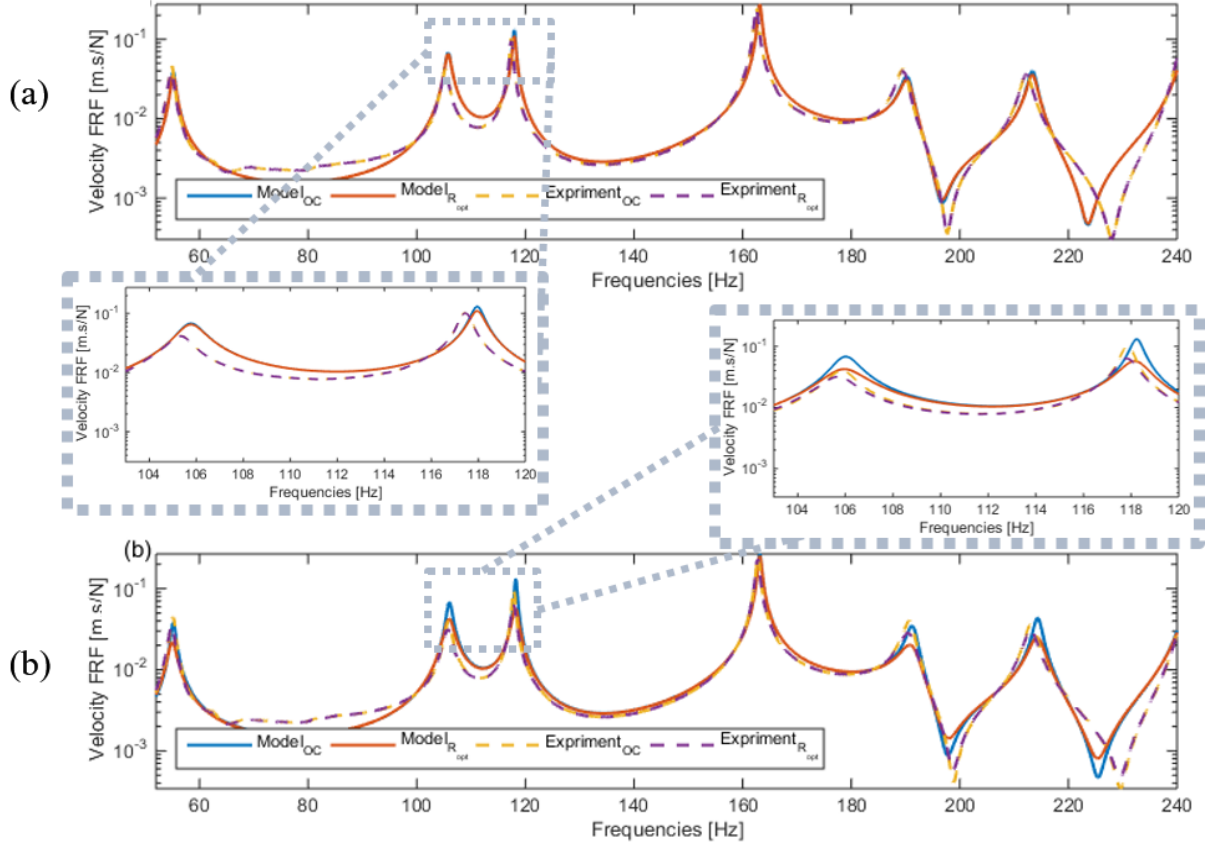

**Figure 3.** Comparison of the analytical model and experimental results (a) connected configuration, (b) separated configuration. The zoomed area shows the reductions in the 2nd and 3rd modes.

It can be observed from the figure that the analytical model accurately predicts the velocity FRFs near resonance and anti-resonance frequencies up to 240 Hz. The velocity percent reductions for each mode are listed in Table 2. As can be seen from the listed results, separated configuration shunt damping performance results are significantly better than the connected configuration.

**Table 2. Velocity percent reduction comparison between the experimental and analytical model**

|  | Connected patches | | Separated patches | |
| --- | --- | --- | --- | --- |
| Mode # | Model | Experimental | Model | Experimental |
| 1 | 16.2 | 15.3 | 43.8 | 41.2 |
| 2 | 1.65 | 1.54 | 36.4 | 35.1 |
| 3 | 5.5 | 5.2 | 44.1 | 43.2 |

## 5.2 Comparison of the analytical and finite-element models

Validation of the analytical model via FEM for a wide range of piezoelectric patch size/thickness is presented in Figure 4 for a single patch. The area ratio is changed from $\frac{A_p}{A_s} = .005$ up to $\frac{A_p}{A_s} = .75$ (whereas $A_p$ and $A_s$ are the piezoelectric patch and the host structure area, respectively). The thickness ratio is also changed from $\frac{h_p}{h_s} = .1$ up to $\frac{h_p}{h_s} = 1$ (whereas $h_p$ and $h_s$ are thickness of the piezoelectric patch and the host structure, respectively. ). Figure 4 shows the comparison of the natural frequencies of the first 6 modes when size and thickness ratios are varied. The first and second column show the Ansys and analytical model results, respectively. The third column shows the percent difference between those results and it can be observed that for all the first 6 modes of the structure, the difference between the analytical and Ansys model remains within 1%. It shows that the analytical model accurately predicts the natural frequencies of the system when the patch size and thickness are varied.

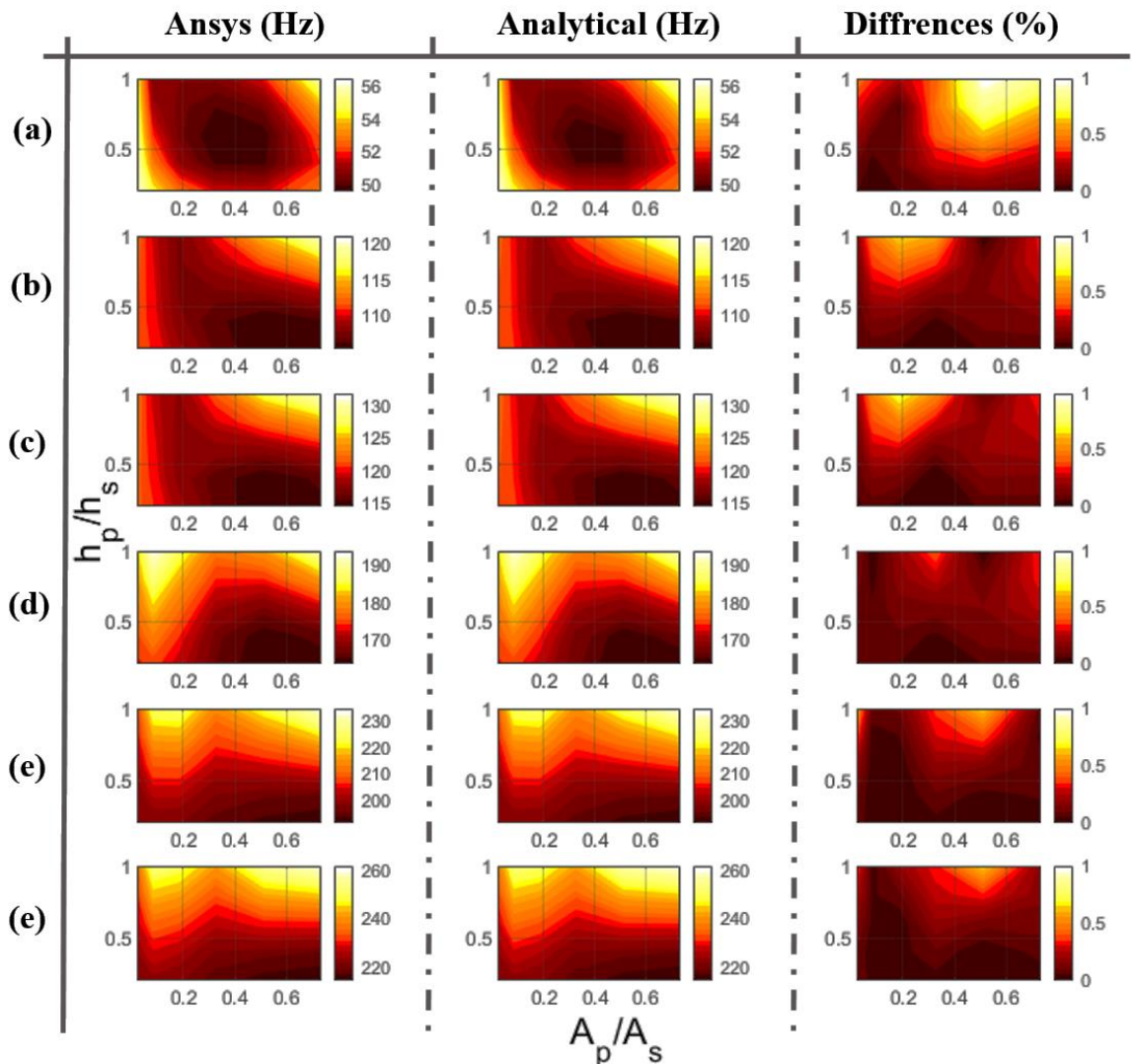

**Figure 4. Comparison of the natural frequencies of the first 6 mode shapes. The first and second column shows the Ansys and analytical model results, respectively. The third column shows the percent difference between those results.**

# 6  Conclusion

This paper presents a methodology and a formulation for separately shunted piezoelectric patches for achieving higher performance on vibration attenuation. The Rayleigh-Ritz method was used for performing the modal analysis of the electro-mechanical system. The developed model includes mass and stiffness contribution of the piezoelectric patches as well as the electromechanical coupling effect. The vibration performance of the separately shunted piezoelectric patches were compared with the connected configurations. An experimental setup was also built to validate the performance of the separately shunted piezoelectric patches. Finite element simulations were performed in ANSYS and compared with the analytical model results for verification of the presented methodology. It was shown that separately shunted piezoelectric patches were more effective compared to connected patches and also more effective when broadband vibration attenuation was considered.